# Current-induced vortex-vortex switching in a nanopillar comprising two Co nano-rings

T. Yang [a)], A. Hirohata, M. Hara

*Frontier Research System, RIKEN, 2-1 Hirosawa, Wako, Saitama 351-0198, Japan*

T. Kimura and Y. Otani

*Institute for Solid State Physics, University of Tokyo, Kashiwa, Chiba 227-8581, Japan,*

*and Frontier Research System, RIKEN, 2-1 Hirosawa, Wako, Saitama 351-0198, Japan*

**Abstract**

We fabricated a current-perpendicular-to-plane pseudo-spin-valve nanopillar comprising a thick and a thin Co rings with deep submicron lateral sizes. The dc current can effectively induce the flux-closure vortex states in the rings with desired chiralities. Abrupt transitions between the vortex states are also realized by the dc current and detected with the giant magnetoresistance effect. Both Oersted field and spin-transfer torque are found important to the magnetic transitions, but the former is dominant. They can be designed to cooperate with each other in the vortex-to-vortex transitions by carefully setting the chirality of the vortex state in the thick Co ring.

Nanomagnets have attracted considerable attention in recent years, because of both the interesting physics at reduced dimensions and the possible industrial applications.[1-3] Extensive studies have been carried out on such nanomagnets as magnetic discs, wires, bars, and rings. Among them, the magnetic nanoring is particularly interesting with the very simple and well-defined magnetic states due to their high symmetry, i.e. the flux-closure "vortex" state and the bidomain "onion" state.[4-7] It is not only well suited for the investigation of fundamental magnetic properties like nucleation, movement and annihilation of domain walls, but also superior in the application because the flux-closure vortex state could prevent the interaction between very close elements.

The magnetic switching in rings from submicron to micron scales has been investigated with magneto-optical Kerr effect (MOKE),[5, 8, 9] magnetic force microscopy (MFM),[6] photoemission electron microscopy (PEEM),[9] and Lorentz microscopy.[10] Magnetoresistance (MR) measurements have also been used to characterize a single ring in the current-in-plane (CIP) configuration.[7, 11] Unfortunately, most of these measurements are not suitable for characterizing an individual tiny magnetic ring element with deep submicron lateral size, which may present unique magnetic behaviors and is requested by such device applications as data storage to achieve high density. In addition, these measurements are usually performed in a uniform linear external field to induce the magnetic transitions. Such a linear magnetic field is not very effective to precisely control the transitions between nonuniform magnetic states, especially the technologically very interesting vortex-vortex transition. It is also difficult to detect the chirality of the vortex state in these measurements, i.e. the closed magnetic flux is clockwise or counterclockwise. On the contrary, the giant magnetoresistance (GMR) measurement in the current-perpendicular-to-plane (CPP) configuration is suitable for characterizing a very small magnetic element and also easy to implement in a real magnetoelectronic device. More attractive is that a current flowing through the structure perpendicularly produces a circumferential Oersted field in the ring, facilitating the formation of flux-closure vortex states and the transition between the vortex states. Such a structure comprising many ring-shaped CoFe layers has been recently experimentally demonstrated with the lateral size of about 600 nm.[12]

In this letter, we report our results on a CPP pseudo-spin-valve (PSV) nanopillar structure comprising two Co rings about 300 nm in outer diameter. It is observed not only the Oersted field but also the spin-transfer torque [13-17] contributes to the vortex-vortex magnetic transitions. The spin-transfer torque has recently been used to assist the domain wall motion in rings in the CIP configuration.[18, 19]

The CPP-PSV nanopillar is fabricated from a magnetic multilayer (bottom)Cu(40nm)/Co2(3nm)/Cu(10nm)/Co1(10nm)/Au(10nm)(top), in which the Co2(3nm)/Cu(10nm)/Co1(10nm)/Au(10nm) layers are patterned into the ring shape with outer/inner diameters being 290 nm/90 nm, as shown in Fig.1 (a). The fabrication process is similar to that for an elliptical nanopillar.[20] The switching of the Co rings is thus studied with the CPP-GMR by applying either an in-plane external field or a vertical dc current. The resistance is measured with a small ac current and the lock-in technique at room temperature. The electrical current flowing from the bottom to the top is defined as positive, as depicted in Fig.1 (a).

Since each Co ring could be in either the vortex state or the onion state, there are various possible magnetic configurations divided into 3 groups shown in Fig.1 (b), according to the relative orientation between the magnetizations in the two Co rings. The CPP resistance should increase as the configuration is varied from parallel (P) to antiparallel (AP).

The magnetic switches of the rings are firstly studied by sweeping the external field. There are three distinct states in the MR loop plotted in Fig. 2 (a), labeled as A, B and C respectively. According to their resistance levels, A should be an AP state in Fig.1 (b), B should be a mixed (M) state, and C should be a P state. From the minor MR loop in Fig.2 (b), state A is stable at 0 external field. Then starting from state A at 0 external field, a dc current up to 50 mA is applied in either the positive or the negative direction. As shown in Fig.2 (b), for either direction, state A is transformed into a low resistance state after a two-step switching at around 20 mA. The two low resistance states persist when the dc current is further swept to ± 50 mA and then back to 0 mA, where they meet each other. Their resistance values indicate they are P states, labeled as D and E respectively in the figure. Further sweeping the dc current between + 50 mA and – 50 mA, it leads to abrupt single-step transitions between D or E state and an AP state, labeled as G in Fig. 2 (c) and F in Fig.2 (d) respectively. Interesting is that while the AP state is induced by a positive dc current and then switched back to the P state by a negative dc current in (c), the opposite switching directions are observed in (d).

To further characterize these P or AP states, the resistance variation with the external field is investigated for each state.

The resistance variation shown in Fig. 3 (a) for D state is much different from that for C state shown in Fig.2 (a), implying that D and C are different P states. No further transition occurs for the C state when the field is increased up to 3500 Oe, typical of the P-O-O configuration. Remarkable is that a transition to a high resistance state appears for D when the field is increased in both directions, which can only be explained with the P-V-V configuration as follows. With increasing the external field in either direction, one magnetic vortex in the P-V-V configuration is firstly switched to the onion state through domain nucleation followed by domain wall propagation, leading to an M configuration and hence the increasing in the resistance. When the rest vortex is also switched to the onion state by a higher field, the M configuration is transformed to the P-O-O configuration and thus a decreasing in the resistance is observed. The resultant P-O-O state is in fact the State C. D state exactly exhibits such a resistance variation when the external field is increased in the positive direction. Ideally, because of the symmetry of the vortex state, the same resistance variation is expected when the field is increased in the negative direction. However, the observed transition to the M configuration is realized by two switches during increasing the field in the negative directions, possibly attributed to a local pinning site blocking the domain wall motion. E state is also a P-V-V state exhibiting almost the same resistance variation but with opposite field directions, as shown in Fig.3 (b). Since both D and E states are induced by the perpendicularly injected dc current, their chiralities should be commensurate with the circumferential Oersted fields, being clockwise and counterclockwise respectively as drawn in Fig. 3(c), if we observe from the top.

It is noticed that the resistances for E and D are slightly lower than that for C state. That may be because that the two onion states in the C state are not exactly parallel to each other under a nonzero external field.

It can be seen from Fig.3 (a) and (b) that the AP states G and F also transform to the C state after two distinct transitions in both positive and negative field directions. Therefore, they should be AP-V-V states. Another AP state A is apparently different from G or F, in the resistance variation induced by either the field or the dc current. Thus A should be an AP-O-O state.

According to a theoretical calculation,[21] the critical circumferential Oersted fields to switch a Co ring increases with the thickness. Therefore, the transition from a P-V-V state to an AP-V-V state should be realized by the vortex-vortex switching in the Co2 ring. Hence the configurations of F and G states can be determined as shown in Fig.3. It seems that the maximum 50 mA dc current cannot produce a field enough to switch the magnetic vortex in Co1 ring.

Based on the above discussions, the current-induced magnetic switches are illustrated in Fig.3, which also explain the opposite switching current directions in Fig.2 (c) and (d).

It should also be noted that critical switching currents in Fig.2 (c) are much higher than the switching currents in Fig.2 (d). The critical switching current densities are about $6.7 \times 10^{11}$ A/m$^2$ and $3.35 \times 10^{11}$ A/m$^2$ respectively. Such a difference is attributed to the spin-transfer torque. The spin-transfer torque applied on Co2 ring by a negative (positive) dc current favors the AP (P) configuration. Simultaneously, the Oersted field created by negative (positive) dc current favors the clockwise (counterclockwise) vortex state in Co2 ring. Thus when a negative dc current is applied on state E, the spin-transfer torque and the Oersted field help each other to reverse the magnetic vortex in Co2 ring, leading to the E-to-F switching. Such a cooperation also occurs for the F-to-E switching. On the contrary, for both D and G states, the spin torque applied by a dc current always conflicts with the Oersted field. Therefore the switching currents for the switches between D and G are much larger than the switching currents for the switches between E and F. Such a significant spin transfer effect is possibly attributed to the small size and the structure comprising only two Co layers with different thicknesses, similar to the extensively studied elliptical nanopillar.[16,17]

In conclusion, we have fabricated a deep submicron-sized CPP-PSV nanopillar, comprising two Co rings with different thicknesses. Flux-closure vortex states can be effectively formed by the dc current with desired chiralities. In addition, abrupt vortex-vortex transitions can also be realized by the dc current and detected by the CPP-GMR effect, which is significant for technological applications. The transitions are dominated by the Oersted field while the contribution of spin-transfer torque is also found important. By carefully choosing the chirality of the thick Co ring, it is possible to exploit both the spin-transfer torque and the Oersted field to switch the vortex states and hence reduce the switching current.

The authors appreciate the help from Nanoscience Development and Support Team of RIKEN.

**Figure captions**

Figure 1 (a) Schematic structure and SEM images, as well as (b) possible magnetic configurations for the fabricated nanopillar. In Each configuration, all the magnetizations could also be in the opposite directions simultaneously.

Figure 2 (a) Full MR loop, (b) minor MR loop (solid line) and resistance variations when the dc current is swept from the A state in the sequences of 0 mA → 50 mA →0 mA (dotted line) and 0 mA → -50 mA →0 mA (dashed line) respectively, and $R\sim I$ loops when the initial states are (c) D and (d) E.

Figure 3 Resistance variations with the external field for (a) D (solid lines) and G (dotted lines) states, as well as (b) E (solid lines) and F (dotted lines) states. (c) is the schematic illustration for the current-induced magnetization switching processes, where the vortex chiralities are sketched in top view. For state A, all the magnetizations could also be in the opposite directions simultaneously.

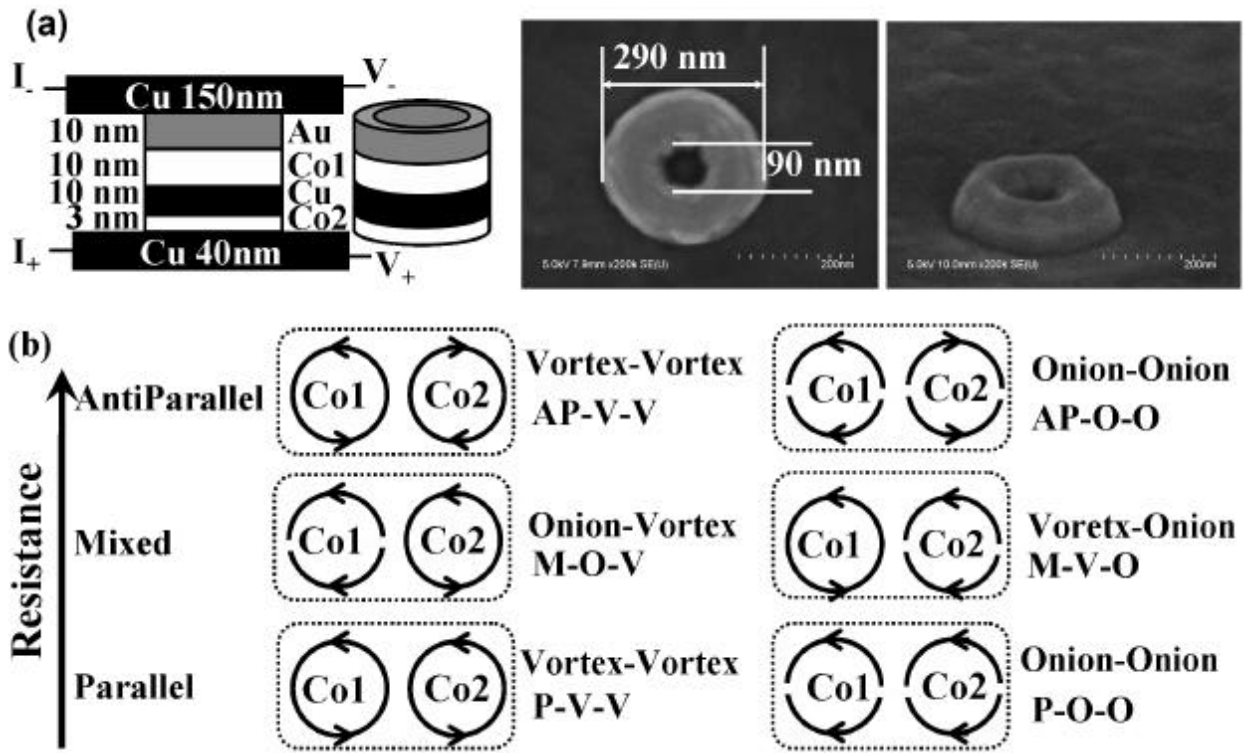

Fig.1

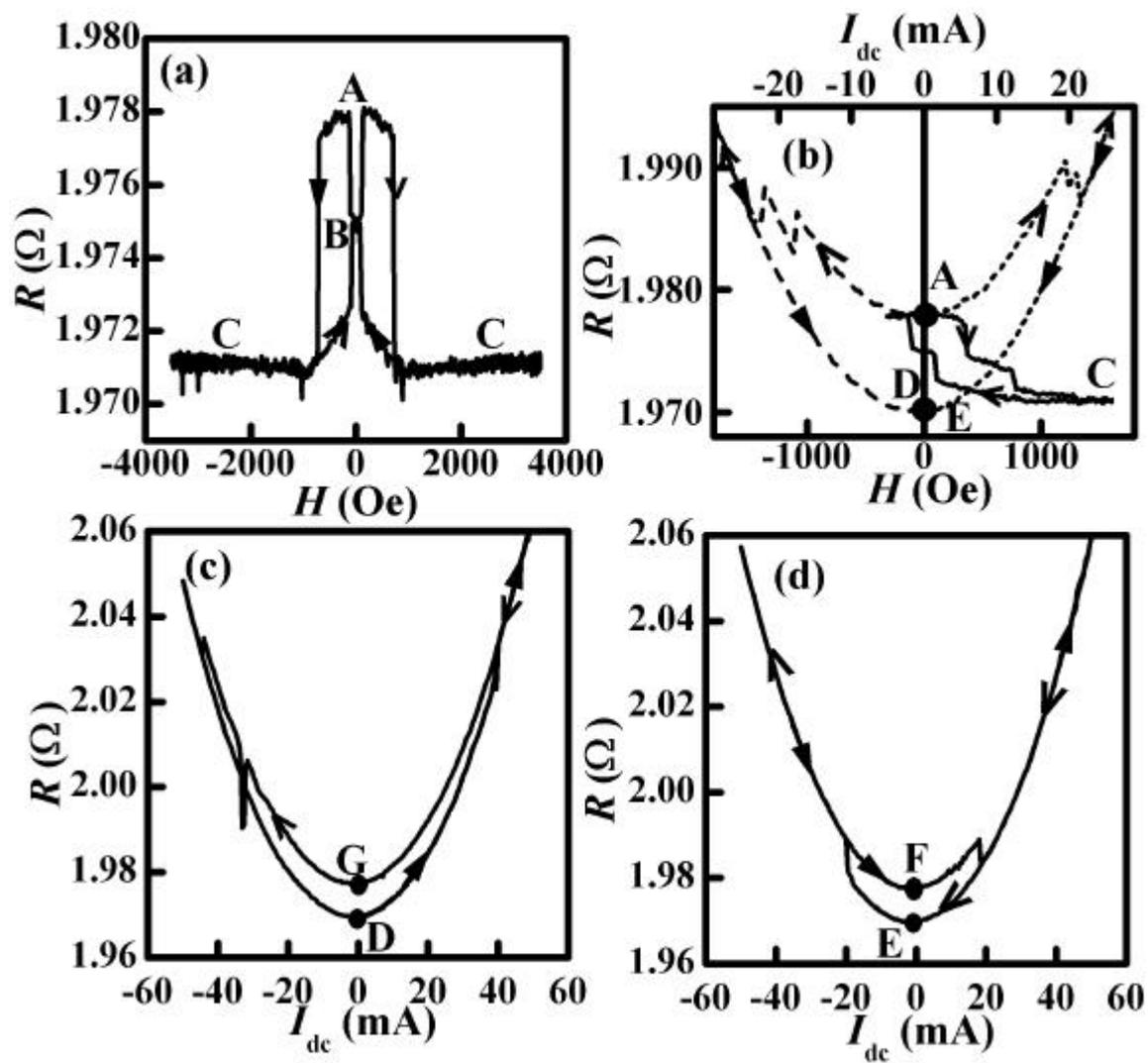

Fig.2

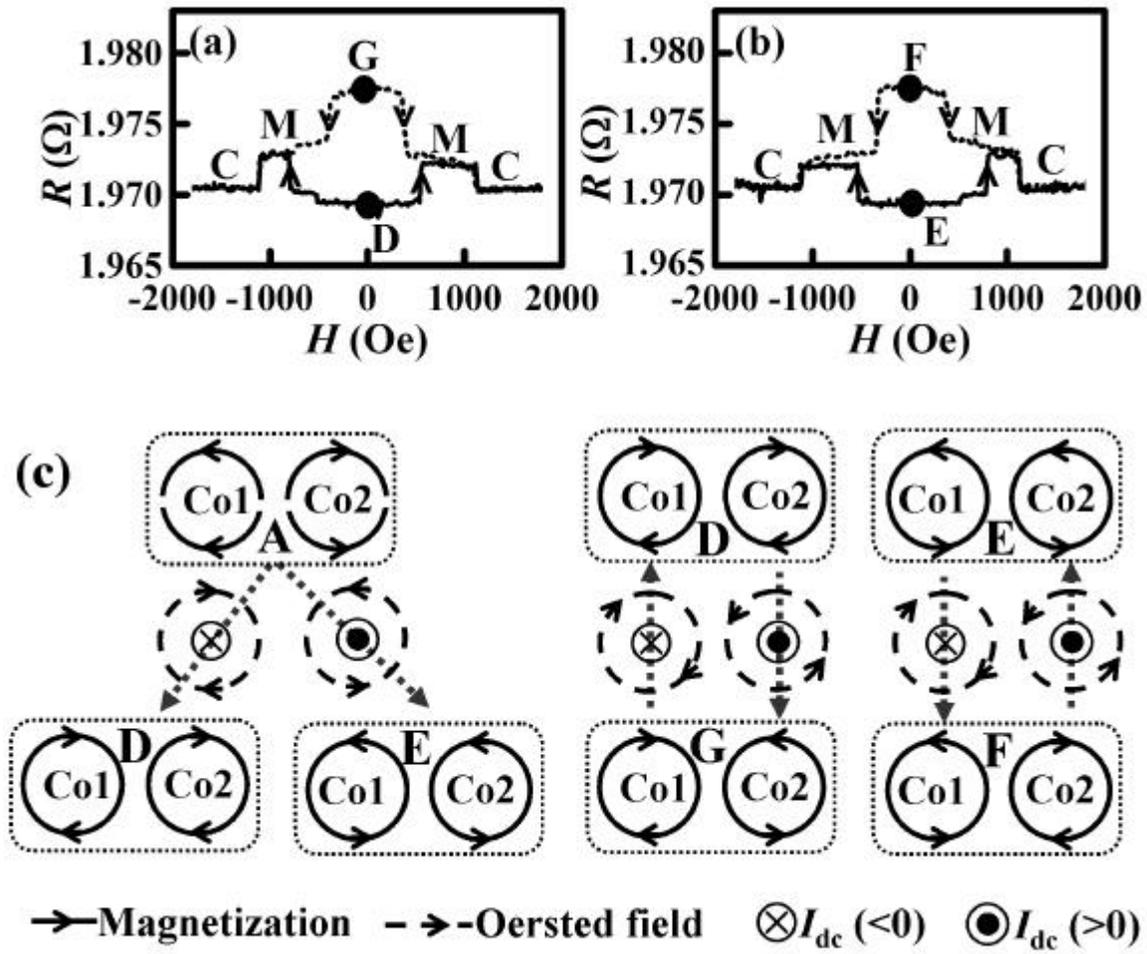

Fig.3